# Numerical and experimental modeling of mixing of impinging jets radially injected into crossflow

**E.V. Kartaev, V.A. Emelkin, M.G. Ktalkherman**
**Khristianovich Institute of Theoretical and Applied Mechanics**
**Siberian Branch of Russian Academy of Sciences**

In some chemical processes, the formation of the counter flow in the colliding-jets regime is the most promising phenomenon if there is a need of fast quenching of an obtained product [1]. Particularly this method can be used to control disperse and phase composition of the final product. The complex calculation and experimental research of the counter collision and mixing of the circular argon jet and aluminum steams at 2000 K, and a relative cold argon jet (1000 K) is presented in [2]. It is shown there that the counter quenching regime enables to control particles grow owing to the flow dilution with the quenching jet and variation of the temperature drop rate.

In view of the above, there is a need to analyze the process of impinging radial jets mixing with the cross flow in the regime of forming axial counter flow. The results of such investigations can be used not only for fast quenching in plasma-chemical reactors, but also for some other technological processes, for example, in the plants of hydrocarbon pyrolysis in a heat-carrier flow [3].

Here we used the following formula for calculating the parameter of depth $h/D$ of penetration of the jets with density $\rho_j$ and flow rate $G_j$, which were radially injected through $n$ orifices of diameter $d$ into a crossflow with density $\rho_m$ and flow rate $G_m$ with the reactor channel of diameter $D$ [4]:

$$\frac{h}{D} = K \frac{D}{dn\mu} \frac{G_j}{G_m} \sqrt{\frac{\rho_m}{\rho_j}}, \qquad (1)$$

where the coefficient $K$ depends on the distance between the orifices, which was taken to be 1.7 in this work, and $\mu$ is the flow rate coefficient of the nozzle, which was taken to be 0.8. The quantity $h$ is the depth of penetration of the radially injected jet into the crossflow at which both flows acquire the same direction. At $h/D < 0.5$, the radially injected jets do not reach the channel axis in the crossflow; at $h/D > 0.5$, the regime of impinging jets is established, which ensures the maximum effectiveness of mixing of the radially injected jets and the crossflow. Obviously, the flows have a turbulent character, and the channel cross section is most uniformly filled by the injected jets.

*Experimental*

Fig. 1 shows the schematic of the flow passage of interest. The heated nitrogen was used as crossflow in the channel. Room-temperature air was used as an injected gas. The air was injected into the channel from the general collector through 8 holes of 5 mm in diameter perpendicularly to the main flow. Upstream from the injection plane, 5 thermocouples were installed (see Fig. 1) to measure the temperature on the channel axis. Each thermocouple was turned about the previous one to the angle of $90°$ in order to decrease the disturbance imposed by them in the flow.

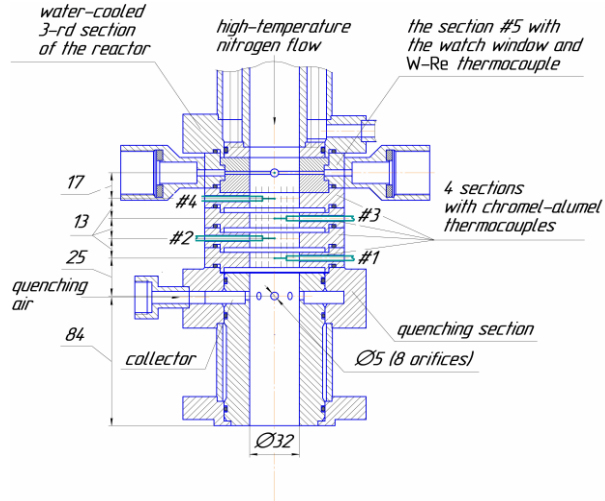

Fig. 1. Schematic of the flow passage: sections 1–5, and quenching section. All dimensions - in mm.

In the experiment, the air was not firstly injected, then its flow rate varied from 4 to 8 g/s, the pitch was 1 g/s. The mass-average velocity of air through each hole ranges from $v_{j1} = 21.2$ m/s (the flow rate of 4 g/s) to $v_{j2} = 42.4$ m/s (the flow rate of 8 g/s). Flow rate of crossflow nitrogen was 0.9 g/s. Mean-mass nitrogen temperature at the anode outlet cross-section was equal to 4000 K, whereas at the inlet of section #5 – 1370 K. Mean velocity of the cross flow of nitrogen at the anode outlet cross section was equal to 134 m/s. The average-mass velocity of the main high-temperature flow ahead of the injection cross section reached $v_{m1} = 4.7$ m/s.

Fig.2 presents the temperature distributions upstream from the injection plane tests with various flow rates of the injected air. The zero coordinate of the abscissa axis corresponds to the jets axes position. It is evident that, without injection, the temperature at the channel axis is roughly similar (curve 1). As the injected air flow rate grows, the upstream temperature decreases. At the maximum flow rate of the quenching gas (8 g/s), in the measurement point of the thermocouple #1, the temperature drops down to 750 K (see Fig. 2, curve 6).

The higher the flow rate of the injected air, the higher the altitude of cold air toward the main high-temperature flow.

Further analysis of the measurement results requires the concept of the counter flow depth of penetration $h_v$. *This value conventionally means the altitude of counter flow about the*

*intersection point of the injection channels axes of colliding jets; at this value, the temperature drops minimum to 100 K against the temperature in this point without disturbing flow.*

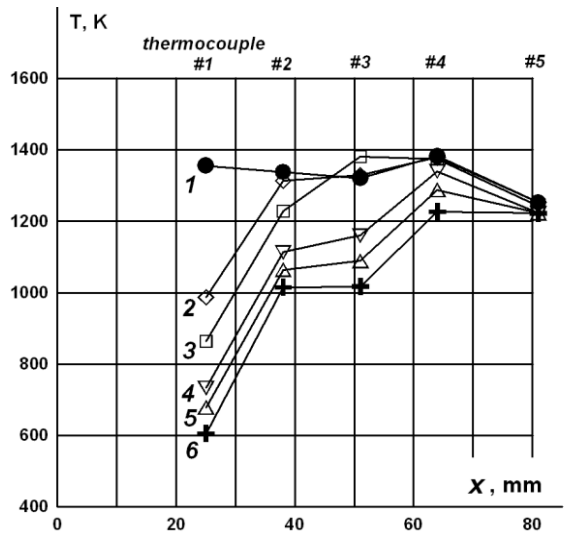

Fig. 2. Temperature profiles on the reactor channel axis, upstream, at the distance *x* from the intersection point of jet channels axes at various air flow rates: 1 – 0 g/sec, 2 – 4 g/sec, 3 – 5 g/sec, 4 – 6 g/sec, 5 – 7 g/sec, 6 – 8 g/sec.

*Simulation*

Let us consider then the simulation results. Fig.3 shows the calculation data for the conditions of the experiment (injected air flow rate $G_j = 8$ g/s).

In the experiment the flow rate of the high-temperature crossflow is minimal (0.9 g/s), whereas the flow rate of the injected gas is maximal.

As is seen from Fig.3a,b, the counter flow altitude upstream the reactor channel axis reaches the value of the channel diameter, the jet velocity toward the main flow right after injected jets collision reaches 20 – 25 m/s. The temperature difference $\Delta T$ on the axis in the undisturbed counter flow area and in the core of the counter quenching jet reaches 1,000 K. It is also seen from Fig.3b,d, that almost complete damping of the radial velocity component takes place in the intersection point of the colliding quenching jets (the $1^{st}$ stagnation point). Fig.3c shows that the turbulent kinetic energy *k* of quenching jets interaction is concentrated in the $1^{st}$ stagnation point area.

One can also see from Fig.3b,d that the stagnation of the quenching jet toward the main counter flow results in the $2^{nd}$ stagnation point forming on the channel axis.

As a result, the most of the high-temperature flow is pushed to the reactor walls and is then captured by the radial jets to the channel axis; a vortex is formed (Fig. 3d). The vortex flow is also observed near the channel wall; it is caused by the main flow passing around the jets which play the role of a "fluid obstacle".

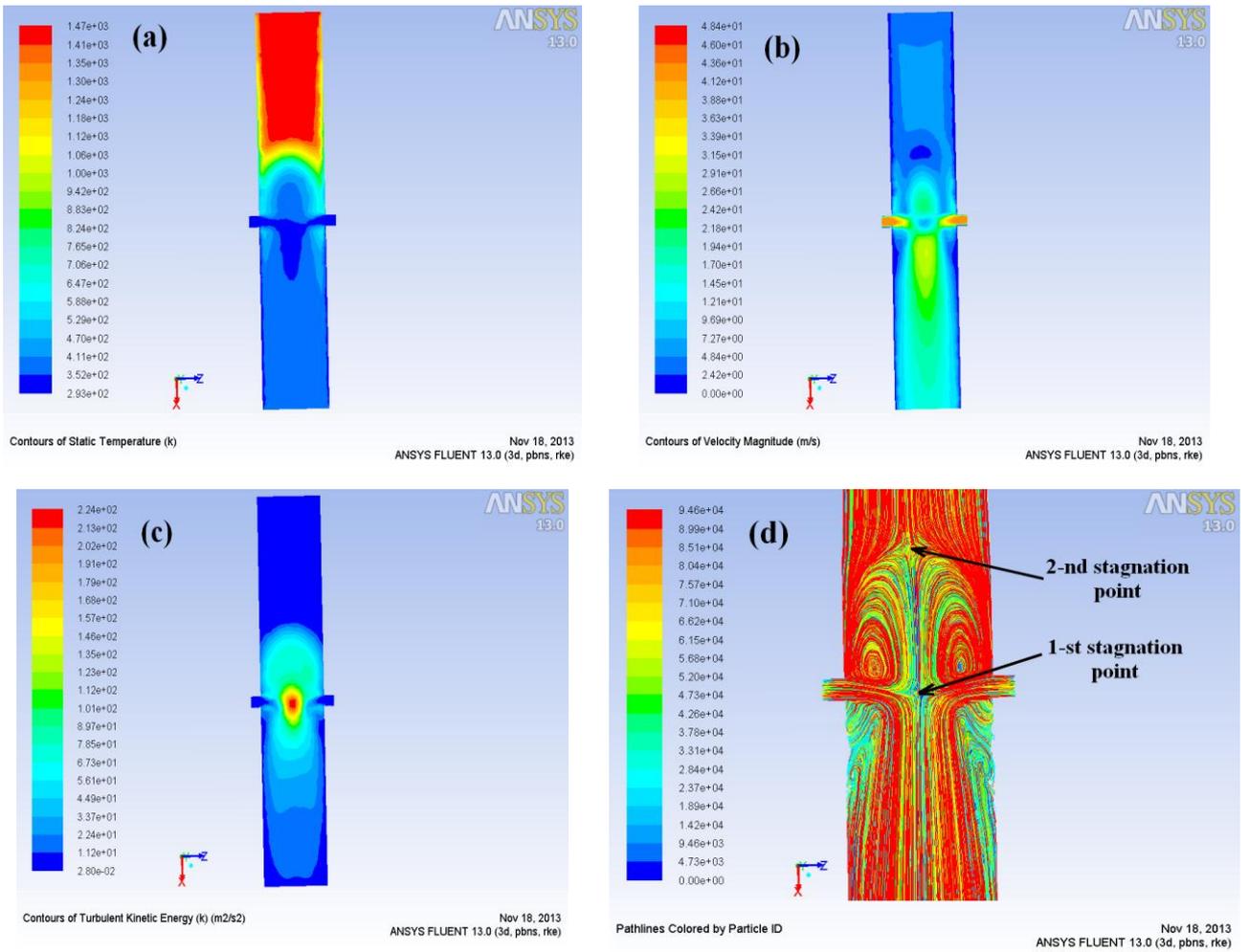

Fig.3. Results of simulation (ANSYS Fluent 13.0) of jets gas-dynamics in the experiment (injected gas flow rate $G_j = 8$ g/s) : (a) – contours of temperature in the reactor; (b) – contours of velocity; (c) – contours of turbulent kinetic energy $k$; (d) – stream lines with two stagnation points on the axis of reactor channel.

*Let us treat $h_v$ as a distance between the 2$^{nd}$ and 1$^{st}$ stagnation points.*

If we accept that the obtained experimental dependence has the same form as formula (1) for the radial penetration, we have a direct connection between the parameter of counter flow jet depth of penetration $h_v/D$ and the parameter of radial penetration $h/D$:

$$h_v / D \approx 0.335 \cdot h / D. \qquad (2)$$

The main flow temperature apparently does not affect the bend of that part of the quenching jets which is formed by the counter flow of the quenching gas, similarly to the bend in the downstream quenching jets.

*Conclusion*

The proposed empirical dependence generalize the results of our measurements of the axial depth of penetration of the counter flow jet which forms in the regime of radial jets collision in the near-axial area of the channel. The results of the numerical calculation illustrate well the flow structure and are in qualitative agreement with the experiment.